# Ru-NMR Studies and Specific Heat Measurements of $Bi_3Ru_3O_{11}$ and $La_4Ru_6O_{19}$

Katsuhiko Tsuchida, Chihiro Kato, Toshiaki Fujita, Yoshiaki Kobayashi
and Masatoshi Sato

*Department of Physics, Division of Material Science, Nagoya University, Furo-cho, Chikusa-ku, Nagoya 464-8602*

**Abstract**

Specific heats measurements and Ru-NMR studies have been carried out for $Bi_3Ru_3O_{11}$ and $La_4Ru_6O_{19}$, which commonly have three-dimensional linkages of edge-sharing pairs of $RuO_6$ octahedra. The Knight shifts, the nuclear spin-lattice relaxation rates $1/T_1$ and the electronic specific heats $C_{el}$ of these systems exhibit anomalous temperature ($T$) dependence at low temperatures, as was pointed out by Khalifah et al. [Nature 411 (2001) 660.] for the latter system based on their experimental data of the resistivity, magnetic susceptibility and electronic specific heat.

Ratios of $1/T_1T$ to the square of the spin component of the isotropic Knight shift, $K_{spin}$ estimated for these systems at low temperatures suggest that they have antiferromagnetic (AF) spin fluctuations. It is confirmed by the fact that the $T$-dependences of $1/T_1T$ and $C_{el}/T$ of the present systems can be explained by the self-consistent renormalization theory for three dimensional itinerant electron systems with AF spin fluctuations. All these results suggest that the AF fluctuations are the primary origin of the characteristics of their low temperature physical behavior.

corresponding author: M. Sato (email: e43247a@nucc.cc.nagoya-u.ac.jp)

**Introduction**

$R_3Ru_3O_{11}$(R=La and Bi) and $La_4Ru_6O_{19}$ with cubic structures have three-dimensional (3D) linkages of edge-sharing pairs of $RuO_6$ octahedra, as shown schematically in Figs. 1(a) and 1(b), respectively.[1,2] They are metallic [3-5] and Ru ions in these systems have an average valency of 4.33. Khalifah et al. first studied $La_3Ru_3O_{11}$ and $La_4Ru_6O_{19}$ by measuring the resistivities ρ, magnetic susceptibilities χ and specific heats, and pointed out that the latter system exhibits non-Fermi liquid behavior at low temperatures, while the former does not.[3,4] For $Bi_3Ru_3O_{11}$, data of transport and magnetic measurements have been presented by the present authors' group in ref. 5 together with those of $Bi_3Os_3O_{11}$, whose structural characteristics are similar to those of $Bi_3Ru_3O_{11}$. Anomalous temperature($T$) dependence of the electronic specific heat $C_{el}$ has also been suggested there.

In order to investigate the origin of the anomalous behavior observed at low temperatures, we have measured the longitudinal relaxation rates $1/T_1$ and the Knight shifts $K$ of $^{99,101}$Ru nuclear moments for $Bi_3Ru_3O_{11}$ and $La_4Ru_6O_{19}$. We have also carried out measurements of specific heats of these systems and deduced the electronic components $C_{el}$. Results of these measurements indicate that antiferromagnetic (AF) fluctuations cause the anomalous $T$ dependent behavior observed in these systems.

**Experiments**

Polycrystalline samples of $Bi_3Ru_3O_{11}$ were prepared by the following method. Mixtures of $Bi_2O_3$ and $RuO_2$ with proper molar ratio were ground and pressed into pellets. These pellets were put in an Au tube, heated for three days at 790 °C and cooled in the furnace. They were reground and pelletized again and heated in the Au tube at 790 °C for 24 h and cooled in the furnace. Polycrystalline samples of $La_4Ru_6O_{19}$ were prepared as follows. $La_2O_3$ dried at 900 °C for 24 h was mixed with $RuO_2$ with proper molar ratio, ground and pressed into pellets. The pellets were sealed in quartz tubes with $KClO_4$ and KCl, heated at 900 °C for three days and then cooled in the furnace. They were washed by distilled water. The samples of these systems were found to be the single phase by X-ray diffraction.

The magnetic susceptibilities χ were measured by using a Quantum Design SQUID magnetometer in the temperature range between 2 K and 300 K. The specific heats $C$ were measured by a thermal relaxation method in the temperature range between 2.3 K and 300 K and by an adiabatic method in the temperature range between 0.8 K and 2.3 K. The $^{99,101}$Ru NMR measurements were carried out by a π/2-τ-π phase–cycled pulse sequence with fixed τ. The NMR spectra were obtained by recording the spin echo intensities with the magnetic field changed step by step. The nuclear spin-lattice relaxation rate $1/T_1$ of $^{99,101}$Ru nuclei was measured by taking the intensity $M(t)$ as a function of time $t$ elapsed after the saturation pulses were applied.

**Results and discussion**

The electronic specific heat $C_{el}$ of $Bi_3Ru_3O_{11}$ obtained by subtracting the phonon contribution $C_{ph}$ from the total specific heat $C_{tot}$ is shown in the form of $C_{el}/T$-$T$ in Fig. 2. It should be noted here that the estimation of $C_{ph}$ is not simple: Because as shown in the inset of Fig. 2 for $Bi_3Ru_3O_{11}$ and $Bi_3Os_3O_{11}$, the $C_{tot}/T$-$T^2$ curve deviates from the straight line, that is, the curve cannot be described by the ordinary relation $C_{tot}/T=\gamma+\beta T^2$ even in the low $T$ region, we cannot simply divide $C_{tot}$ into two terms, $C_{el}$ and $C_{ph}$.[5] A possible origin of the deviation is that the Debye temperatures of these systems are very small, because they have heavy Bi ions. Then, we just tried to fit a polynomial form up to the third order of $T$, to the observed data of $C_{tot}/T$ of $Bi_3Os_3O_{11}$. The result of the fitting is shown in the inset of Fig. 2 by the solid line, where the electronic part $C_{el}/T=\gamma$ ($\sim 17$ mJ/mol Os K$^2$) and the sum of other terms ($=C_{ph}/T$) are separated. For the specific heat $C_{tot}$ of $Bi_3Ru_3O_{11}$, we realize that the broken line obtained by a vertical shift of the solid line can well reproduce the result of $Bi_3Ru_3O_{11}$ above 6 K. Based on this result, we introduce an assumption that the phonon contributions of both systems are equal, which enables us to deduce $C_{el}$ of $Bi_3Ru_3O_{11}$. The result is shown in Fig. 2 in the form of $C_{el}/T$-$T$.

The $T$ dependence of $C_{el}$ of $Bi_3Ru_3O_{11}$ reminds us the anomalous behavior of the electronic specific heat found for $La_4Ru_6O_{19}$, for which the anomalous $T$-linear dependence of the resistivity $\rho$ at low temperatures has also been reported. For the present $Bi_3Ru_3O_{11}$, the $T$-dependence of the resistivity $\rho$ is roughly described by the relation $\rho = \rho_0 + AT^\varepsilon$ with $\varepsilon \cong 1.8$ in the low temperature region (0.3 K$\leq T \leq$15 K),[5] which is close to the value 2.0 expected for Fermi liquid. Because samples of $La_4Ru_6O_{19}$ were obtained in the form of very small crystallites, we could not measure the resistivity $\rho$.

Figure 3 shows the $T$-dependence of the magnetic susceptibilities $\chi$ taken for the present samples of $Bi_3Ru_3O_{11}$[5] and $La_4Ru_6O_{19}$. The result for $La_4Ru_6O_{19}$ is consistent with the data reported previously.[3]

$^{99,101}$Ru-NMR spectra of $Bi_3Ru_3O_{11}$ and $La_4Ru_6O_{19}$ are shown in Figs. 4(a) and 4(b), respectively. The powder pattern calculated for the former system by using the asymmetric parameter $\eta \sim 0.5 \pm 0.05$, the nuclear electronic quadrupole ($eqQ$) frequencies $^{99}\nu_Q \sim 2.37 \pm 0.05$ MHz and $^{101}\nu_Q \sim (^{101}Q/^{99}Q) \times ^{99}\nu_Q = 5.8\,^{99}\nu_Q$ and the Knight shifts $^{99,101}K_x = ^{99,101}K_y = ^{99,101}K_z \sim -0.3 \pm 0.05$ % reproduce the data as shown by the solid line in Fig.

4(a). The powder pattern calculated for the latter system by using $\eta \sim 0.85 \pm 0.1$, $^{99}\nu_Q \sim 2.7 \pm 0.1$ MHz, $^{101}\nu_Q \sim 5.8\,^{99}\nu_Q$, $^{99,101}K_x \sim -1 \pm 0.1$ % and $^{99,101}K_y = {}^{99,101}K_z \sim -3 \pm 0.1$ % reproduce the data as shown by the solid line in Fig. 4(b). The peaks of the $^{101}$Ru spectra of these systems were not detectable, due to the broadening caused by the large $eqQ$ interaction, as in the case of $RuO_2$.[6]

Figure 5 shows the $T$-dependence of the isotropic part of Knight shifts, $K$ ($\equiv (K_x+K_y+K_z)/3$) estimated from the spectra taken under the magnetic fields $H \sim 10$ T for $Bi_3Ru_3O_{11}$ and $La_4Ru_6O_{19}$. The shift $K$ of $Bi_3Ru_3O_{11}$ is negative and almost $T$-independent in the temperature range between 1.8 K and 20 K. For $La_4Ru_6O_{19}$, the absolute value of $K$ increases with decreasing $T$ in the same temperature range. These results are consistent with the observed behavior of their magnetic susceptibilities $\chi$ shown in the inset of Fig. 3. From the $K$-$\chi$ plot (shown in the inset of Fig.5), the hyperfine coupling constant of the spin part, $A_{spin}$ is estimated to be $\sim -185 \pm 20$ kOe/$\mu_B$ for $La_4Ru_6O_{19}$, which is nearly equal to the value reported for $CaRuO_3$,[6] although their average valences are different. It is difficult to estimate the value of $A_{spin}$ of $Bi_3Ru_3O_{11}$, because $K$ and $\chi$ are almost $T$-independent.

The nuclear-spin relaxation rates $1/T_1$ of $Bi_3Ru_3O_{11}$ and $La_4Ru_6O_{19}$ have been measured at the positions indicated by the arrows in Figs. 4(a) and 4(b), respectively. If the measurements are carried out under a condition that the central transition is saturated but the populations in the other levels are not affected, relaxation curves can be described

$$\frac{M(\infty) - M(t)}{M(\infty)} = (1 - \lambda - \mu)\exp(-\frac{t}{T_1}) + \lambda\exp(-\frac{6t}{T_1}) + \mu\exp(-\frac{15t}{T_1}) \qquad (1)$$

with $\mu=0.794$ and $\lambda=0.178$. However, because there are not only the contribution of central transition line but also the weak contribution of the satellite transition lines to the spectral intensities at the measuring positions, the observed relaxation curves do not precisely agree with the theoretical curve. Here, an expression with slightly modified values $\mu=0.77$ and $\lambda=0.17$ have been used to reproduce the observed decay curves or to consider *effectively* possible effects of the actual experimental situation just stated above. A typical example of the relaxation curves taken for $La_4Ru_6O_{19}$ is shown in the inset of Fig. 6. The obtained data of $1/T_1T$ are plotted against $T$ in Figs. 6 and 7 for $La_4Ru_6O_{19}$ and $Bi_3Ru_3O_{11}$, respectively. The values increase with decreasing $T$ for both systems of $Bi_3Ru_3O_{11}$ and $La_4Ru_6O_{19}$ in the studied temperature range, indicating that the simple Korringa relation does not hold in the present systems.

Here, to describe the spin contribution to the relaxation rate divided by $T$, $(1/T_1T)_{spin}$,

we simply use a modified Korringa relation

$$\left(\frac{1}{T_1T}\right)_{\text{spin}} = \frac{8\pi^2 k_B}{h}\left(\frac{\gamma_n}{\gamma_e}\right)^2 K_{\text{spin}}^2 K(\alpha) \qquad (2)$$

usually used for a single band electron system with strong correlation, where $K_{\text{spin}}$ is the 4$d$-spin contribution to the isotropic component $K$ of the Knight shift and $\gamma_e$ and $\gamma_n$ are the gyromagnetic ratios of electrons and Ru nuclei, respectively. If ferromagnetic spin fluctuations are dominant in the electron systems, $K(\alpha)$ is expected to be smaller than unity, while if antiferromagnetic fluctuations are dominant, $K(\alpha) > 1$ holds.

In the actual estimation of $K_{\text{spin}}$ of $La_4Ru_6O_{19}$, by using the orbital hyperfine coupling constant, $A_{\text{orb}} \sim 385$ kOe/$\mu_B$ estimated for free $Ru^{4+}$ ions,[6] the orbital contribution of the Knight shift $K_{\text{orb}}$ and the susceptibility $\chi_{\text{orb}}$ were determined to be $0.39\pm0.15$ % and $(5.6\pm2.0)\times10^{-5}$ emu/mol-Ru, respectively. For $Bi_3Ru_3O_{11}$, $A_{\text{spin}}$ was assumed to be equal to that of $La_4Ru_6O_{19}$, and $K_{\text{orb}}$ and $\chi_{\text{orb}}$ were determined to be $0.82\pm0.08$ % and $(1.21\pm0.10)\times10^{-4}$ emu/mol-Ru, respectively. We neglected the orbital contributions to the nuclear spin-lattice relaxation rates divided temperature $(1/T_1T)_{\text{orb}}$, because $(1/T_1T)_{\text{orb}} \sim 0.2$ s$^{-1}$K$^{-1}$ in other Ru oxides[6] is much smaller than the values $1/T_1T$ (>2 s$^{-1}$K$^{-1}$ for $Bi_3Ru_3O_{11}$ and >20 s$^{-1}$K$^{-1}$ for $La_4Ru_6O_{19}$) in the studied $T$ regions. The values of $K(\alpha)$ increase with decreasing $T$ and at the lowest temperature (~1.8 K) studied here, they are $5.0\pm0.6$ and $5.0\pm0.5$ for $Bi_3Ru_3O_{11}$ and $La_4Ru_6O_{19}$, respectively. These values suggest that the systems have AF spin fluctuations. (Note the following fact. When eq. (2) cannot be properly applied to the present systems for the possible reason that there exists the degeneracy of the $t_{2g}$ orbitals at the Fermi level, the values of $K(\alpha)$ are larger than those estimated here. Then, the present result does not change.)

To study if the present systems actually have strong AF fluctuations, we analyze the data of $1/T_1T$ and $C_{el}/T$ by using the self-consistent renormalization (SCR) theory[7-11] for itinerant electron systems. According to the theory, the $T$-dependence of $1/T_1T$ for 3D systems with AF fluctuations is given by[8]

$$\frac{1}{T_1T} = \frac{3h\gamma_n^2 A_{\text{spin}}^2}{16\pi k_B T_A T_0 \sqrt{y}} \qquad (3)$$

where $T_0$ and $T_A$ specify the frequency and momentum spread of spin fluctuations. And $y$ is the dimensionless inverse staggered susceptibility defined by

$$y = \frac{1}{2k_B T_A \chi(Q)} \qquad (4)$$

where $Q$ is AF wave vector and $\chi(Q)$ is the staggered susceptibility in units of $(2\mu_B)^2$

per unit energy per atom. The $T$-dependence of $C_{el}/T$ for 3D systems with AF fluctuations is given by[9]

$$\frac{C_{el}}{T} = \frac{3k_B N_0}{2T_0}\left(1 - \frac{\pi}{2}\sqrt{y}\right) + \gamma_0 \tag{5}$$

where $N_0$ is the number of Ru atoms and $\gamma_0$ is the electronic specific heat coefficient for non-interacting electrons. Here, we just assume the Curie-Weiss form of $\chi(Q)$ ($=C/(T+\theta)$). Then $y_0$ which indicates the distance from the quantum critical point (QPC) is defined by the value of $y$ at $T=0$ K ($=\theta/2k_B T_A C$). Then, we have found that the observed $T$ dependence of $1/T_1 T$ and $C_{el}/T$ of $Bi_3Ru_3O_{11}$ can be fitted by these equations with common parameters as shown in Figs. 2 and 7. In the fitting, to the data of $C_{el}/T$, $\gamma_0$ of 15.1 mJ/mol-Ru K$^2$ obtained from the experimental $C_{el}/T$ above 6 K has been used. The parameters ($\theta, T_0, T_A, C, y_0$) determined by the fitting are summarized in Table I.

From the specific heat data of $La_4Ru_6O_{19}$ shown in the inset of Fig. 8 in the form of $C_{tot}/T-T^2$, the electronic component of $C_{el}/T$ have been extracted by subtracting the phonon component, which is shown by the straight line. Although the $C_{el}/T$-$T$ curve thus obtained for the present sample is somewhat different from that of the previous report.[3] The $1/T_1 T$-$T$ curve below 20 K and $C_{el}/T$ below 10 K for $La_4Ru_6O_{19}$ can be fitted by the eqs. (3) and (5) with common parameters as shown in Figs. 6 and 8 by the solid lines, respectively. In the fitting to the data of $C_{el}/T$, $\gamma_0$ of 5.1 mJ/mol-Ru K$^2$ estimated from the data of $C_{el}/T$ above 10 K has been used. The parameters ($\theta, T_0, T_A, C, y_0$) determined by the fitting are summarized in Table I. The rather good fittings obtained for both $Bi_3Ru_3O_{11}$ and $La_4Ru_6O_{19}$ indicate the important role of the AF spin fluctuations in the determination of the low temperature physical behaviors of these systems.

We have measured the $H$-dependence of $C(=C_{tot})$, $K$ and $1/T_1 T$ of $Bi_3Ru_3O_{11}$ and $La_4Ru_6O_{19}$ (The data are not shown.) and found that $K$ and $1/T_1 T$ of $Bi_3Ru_3O_{11}$ do not depend on $H$ appreciably at 4 K in the range between 6.5 T and 10 T. For $La_4Ru_6O_{19}$, no appreciable $H$-dependence of $K$ and $1/T_1 T$ have been found either, between 6.78 T and 10 T at 4.7 K and 1.9 K. Although $C/T$ of $La_4Ru_6O_{19}$ reported previously[3] changes from 40 mJ/mol Ru K$^2$ to 45 mJ/mol Ru K$^2$ with changing the field from 0 T to 6 T at 5 K, present $C/T$ data for both $La_4Ru_6O_{19}$ and $Bi_3Ru_3O_{11}$ measured at various fixed fields between 0 T to 7 T in the temperature range from 2.3 K to 300 K do not exhibit significant $H$-dependence. For $Bi_3Ru_3O_{11}$, Lee et al. have also reported that $C/T$ has no significant $H$-dependence.[10] The non-existence of the $H$-dependence of $C$, $K$ and $1/T_1 T$ of the present systems are consistent with the idea that the anomalous behavior of these systems are caused by the AF spin fluctuations, because the effect of the magnetic field is expected not to be strong on the AF spin correlation.

The parameters obtained for $Bi_3Ru_3O_{11}$ and $La_4Ru_6O_{19}$ are considered to be reasonable, judging from the comparison with those of $V_{1.973}O_3$[8] and $Ce_{1-x}La_xRu_2Si_2$[12] which are also considered to have itinerant electrons with AF spin fluctuations (see Table II). It is also known from the $y_0$ values of $Bi_3Ru_3O_{11}$ and $La_4Ru_6O_{19}$ that the latter system is closer to the QPC than the former. It may be an origin of the difference between the $T$-dependences of $\chi$.

It has been suggested that the small distance between the Ru-Ru atoms within a edge-sharing pair of $RuO_6$ octahedra is important for the realization of the non-Fermi liquid behavior.[3, 4] The distance in $La_4Ru_6O_{19}$ which exhibits the non-Fermi liquid behavior is 2.448 Å and it is 2.993 Å for $La_3Ru_3O_{11}$ with ordinary Fermi liquid behavior reported by Khalifah et al.[3] $Bi_3Ru_3O_{11}$ with an intermediate Ru-Ru distance of 2.60 Å[1] also exhibits the $T$-dependent behavior of $C_{el}/T$ and $1/T_1T$. But the exponent ε related to the $T$ dependence of the resistivity of this system is not far from the value expected for ordinary strongly correlated electrons. These observations are consistent with the suggestion.

Now, we have shown that the low temperature physical properties of these systems that AF spin fluctuations cause the anomalous $T$ dependence of $C_{el}/T$ and $1/T_1T$.

**Conclusion**

Ru NMR and other kinds of measurements have been carried out for two metallic systems $Bi_3Ru_3O_{11}$ and $La_4Ru_6O_{19}$. The Knight shifts $K$ for $La_4Ru_6O_{19}$ and the longitudinal relaxation rates divided by $T$, $1/T_1T$ for both systems exhibit significant $T$ dependence at low temperatures. The electronic specific heat coefficient $\gamma=C_{el}/T$ of $Bi_3Ru_3O_{11}$ also exhibits strong $T$ dependence below 6 K. The large values of $K(\alpha)$ obtained for these systems at 1.8 K indicate that there exist strong AF spin fluctuations and these anomalous $T$ dependences can be well understood by the SCR theory, where we have found $La_4Ru_6O_{19}$ is closer to the QPC than $Bi_3Ru_3O_{11}$.

**Acknowledgements**

The present work is supported by Grants-in-Aid for Scientific Research from the Japan Society for the Promotion of Science (JSPS) and by Grants-in-Aid on priority area from the Ministry of Education, Culture, Sports, Science and Technology.

**Figure captions**

Fig. 1  Schematic structures of $R_3Ru_3O_{11}$(a) and $La_4Ru_6O_{19}$(b). $RuO_6$ octahedra are shown by shaded polyhedra (with Ru atoms inside). R atoms are shown by filled circles.

Fig. 2  $T$-dependence of the electronic specific heat divided by $T$, $C_{el}/T=(C_{tot}-C_{ph})/T$ of $Bi_3Ru_3O_{11}$. The solid line shows the result of the fitting below 6 K by eq. (5). The inset shows the specific heat data divided by $T$, $C_{tot}/T$ of $Bi_3Os_3O_{11}$ and $Bi_3Ru_3O_{11}$ against $T^2$, where the solid line shows the polynomial fit to the data of $Bi_3Os_3O_{11}$ above 6 K and the broken line is obtained by adding a constant to the solid line. See the text for details.

Fig. 3  $T$-dependence of the magnetic susceptibilities of $Bi_3Ru_3O_{11}$ and $La_4Ru_6O_{19}$ measured with $H$= 5 T. The inset shows the data below 25 K.

Fig. 4  Ru NMR spectra of $Bi_3Ru_3O_{11}$ at 4.7 K(a) and those of $La_4Ru_6O_{19}$ at 10 K(b). In each figure, the fitted result is shown by the solid line.

Fig. 5  $T$-dependence of the isotropic part of the Knight shifts of $Bi_3Ru_3O_{11}$ and $La_4Ru_6O_{19}$ ($H$~10 T). The inset shows the $K$-$\chi$ plots.

Fig. 6  $1/T_1T$ of $La_4Ru_6O_{19}$ measured with $H$ ~10 T is plotted against $T$. The solid line is the result of the fitting by eq. (3). Inset shows the relaxation curve of $La_4Ru_6O_{19}$ taken under the field of ~10 T at 7 K. The solid line is the result of the fitting by eq. (1).

Fig. 7  $T$-dependence of $1/T_1T$ of $Bi_3Ru_3O_{11}$ taken with $H$ ~10 T. The solid line is the result of the fitting by eq. (3).

Fig. 8  $T$-dependence of the electronic specific heat divided by $T$, $C_{el}/T=(C_{tot}-C_{ph})/T$ of $La_4Ru_6O_{19}$. The solid curve is the result of the fitting by eq. (5). In the inset, the specific heat divided by $T$, $C_{tot}/T$ of $La_4Ru_6O_{19}$ is plotted against $T^2$. The straight line shows the estimation of $C_{ph}$.

Table I. Characteristic parameters of the SCR theory for $Bi_3Ru_3O_{11}$ and $La_4Ru_6O_{19}$ obtained by the fittings to the data of $C_{el}/T$ and $1/T_1T$

|  | $\theta$(K) | $T_0$(K) | $T_A$(K) | $C(10^{14}$ K/erg) | $y_0$ |
|---|---|---|---|---|---|
| $Bi_3Ru_3O_{11}$ | 1.059 | 446 | 169 | 4.53 | 0.05 |
| $La_4Ru_6O_{19}$ | 1.092 | 161 | 55.6 | 20.2 | 0.035 |

Table II. Characteristic parameters of the SCR theory for $V_{1.973}O_3$[8] and $Ce_{1-x}La_xRu_2Si_2$[12]

|  | $T_0$(K) | $T_A$(K) | $y_0$ |
|---|---|---|---|
| $V_{1.973}O_3$ | 390 | 4800 | -0.0035 |
| $CeRu_2Si_2$ | 14.1 | 16 | 0.31 |
| $Ce_{0.95}La_{.005}Ru_2Si_2$ | 14.7 | 14 | 0.1 |
| $Ce_{0.925}La_{.0075}Ru_2Si_2$ | 14.2 | 11 | 0.05 |

(a) 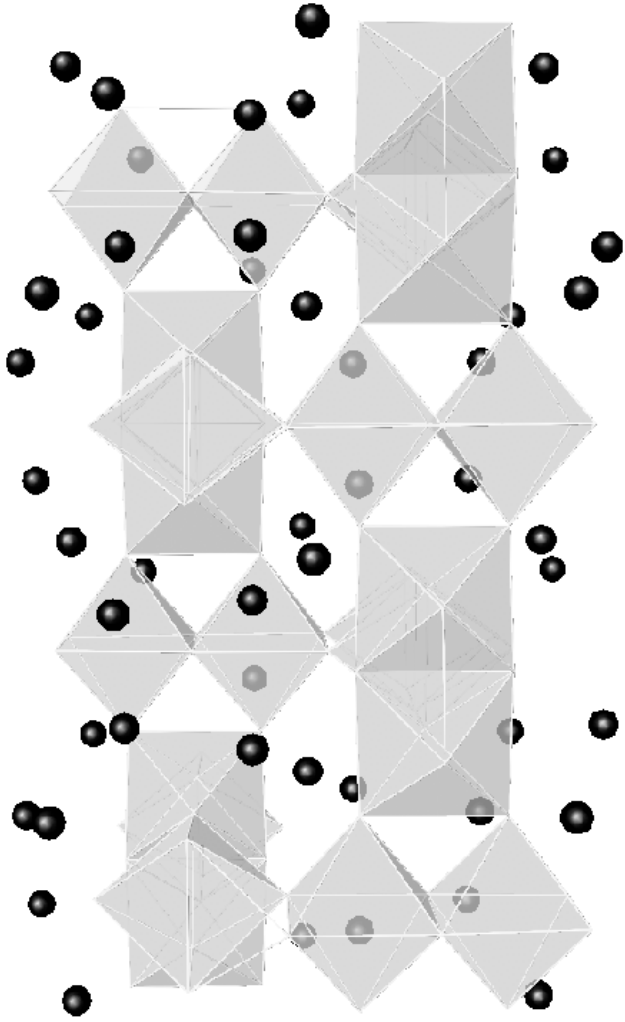  (b) 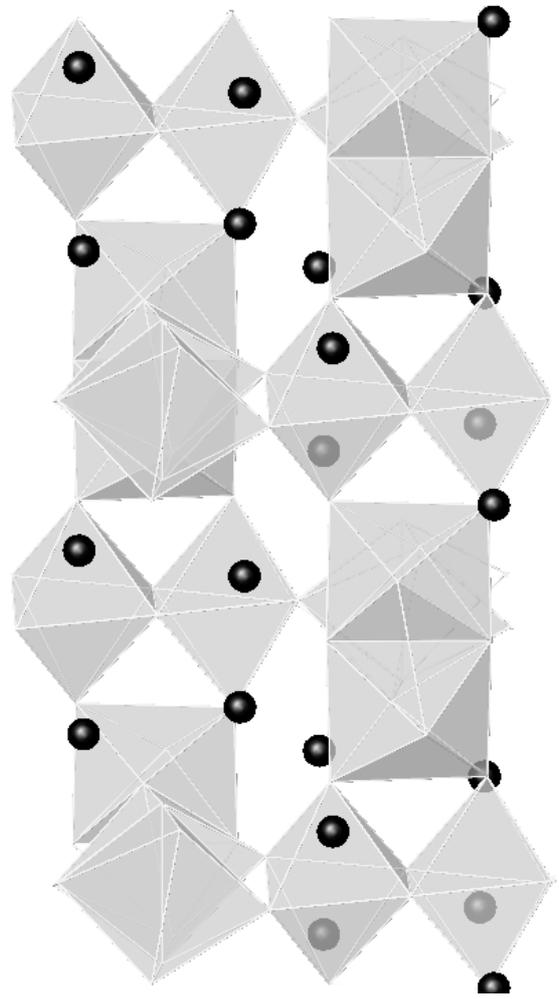

R$_3$Ru$_3$O$_{11}$
(R=La and Bi)

La$_4$Ru$_6$O$_{19}$

Fig. 1
Tsuchida *et al.*

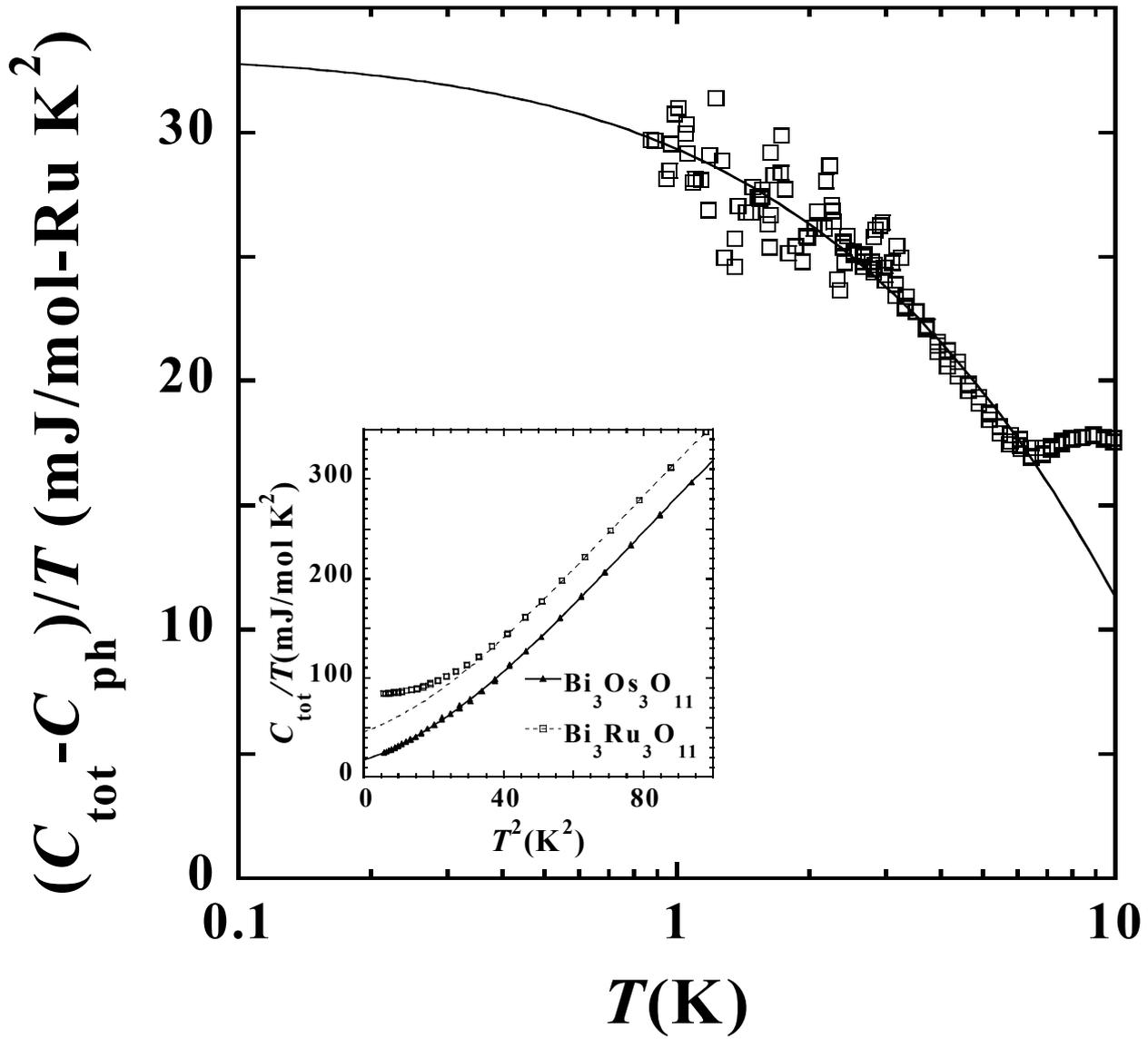

Fig.2
Tsuchida *et al.*

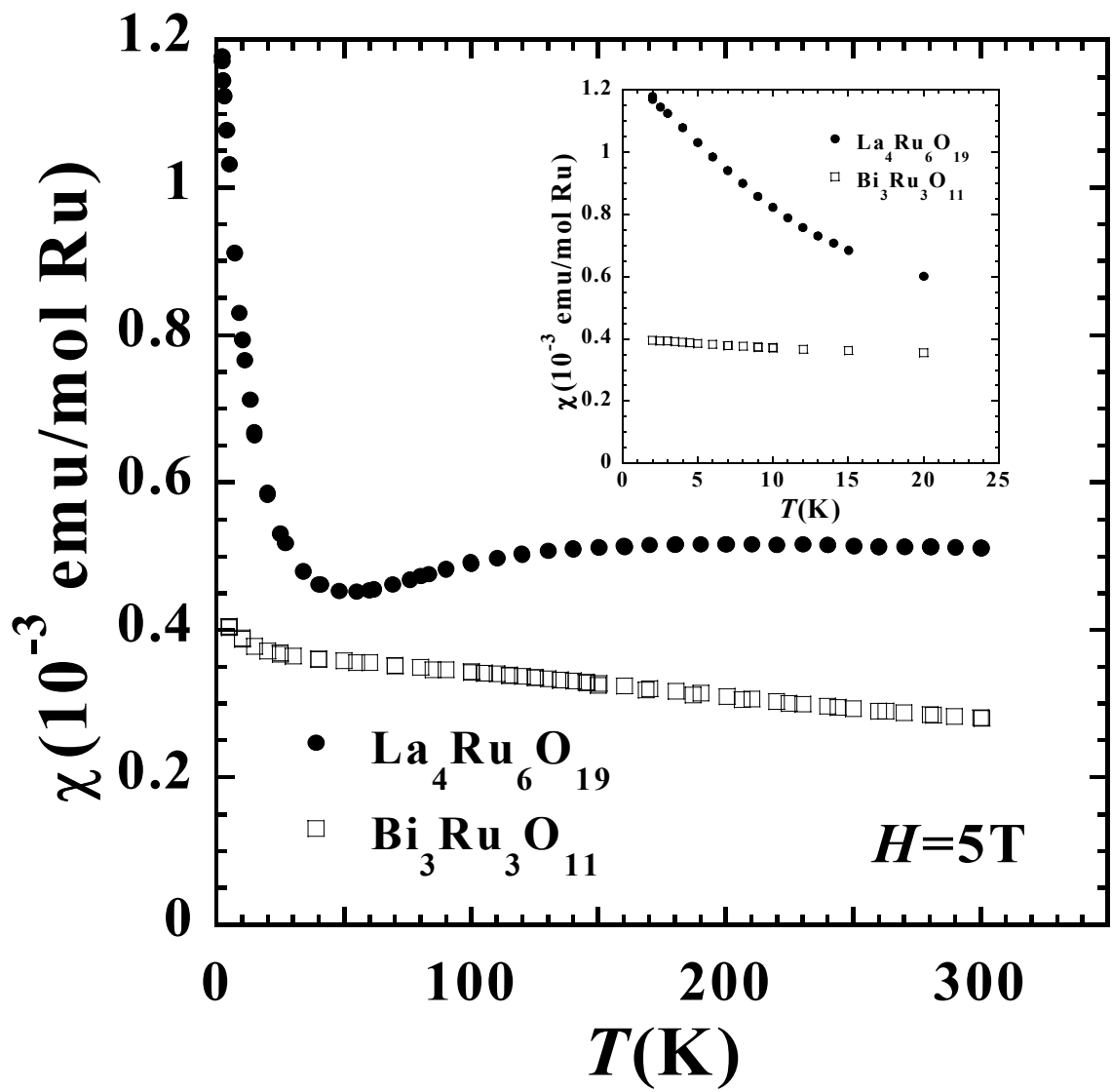

Fig.3
Tsuchida *et al*.

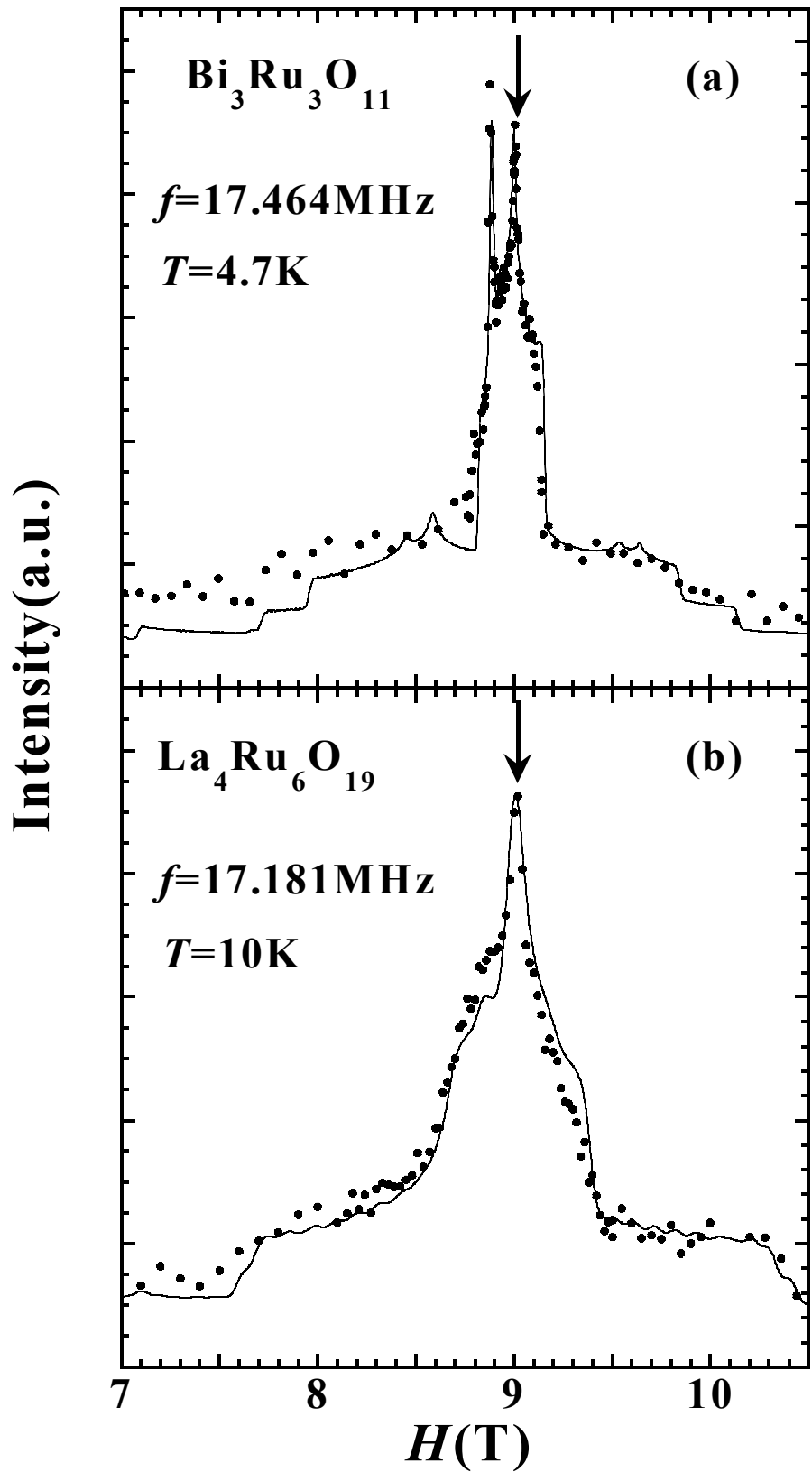

Fig.4
Tsuchida *et al.*

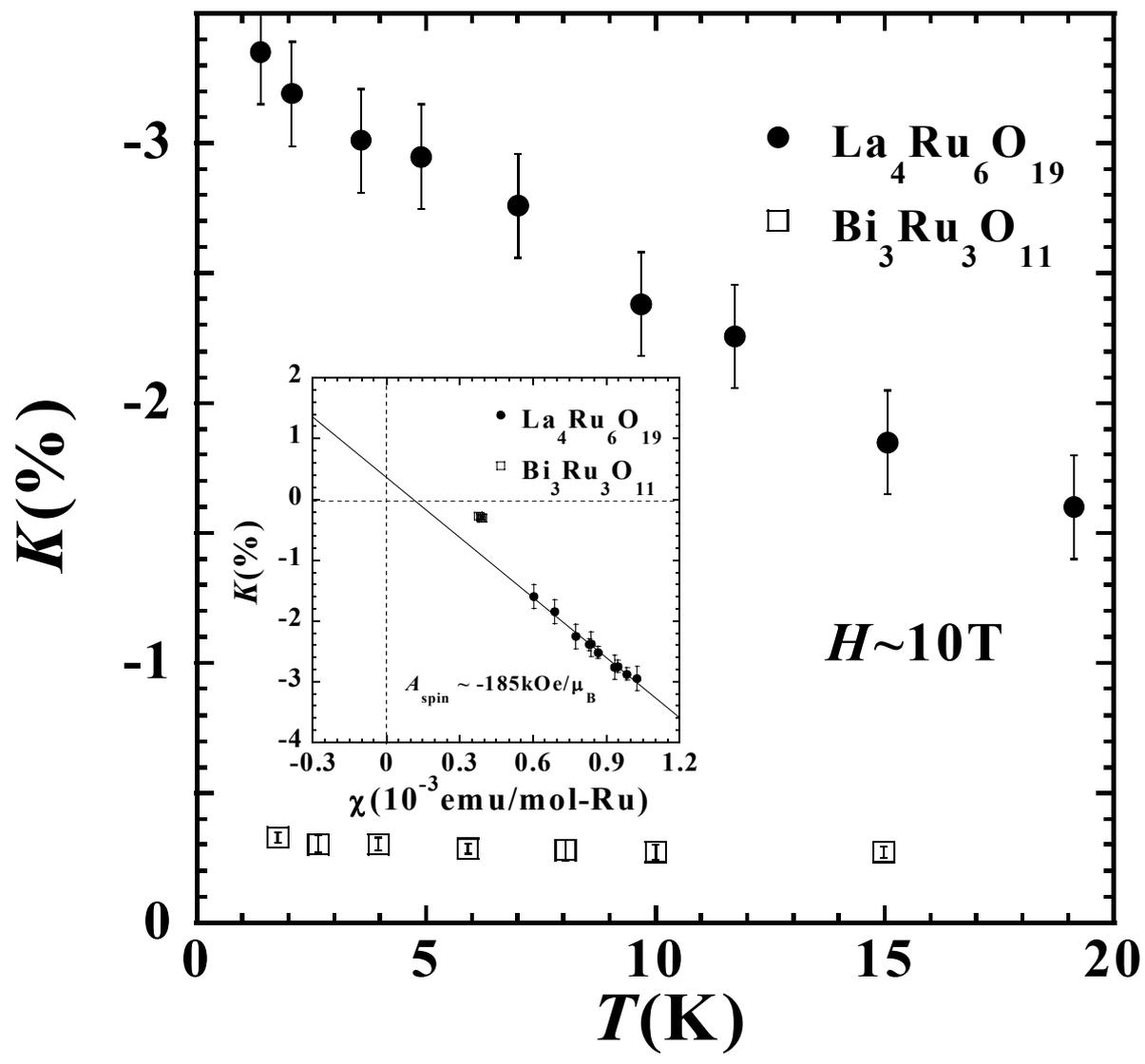

Fig.5
Tsuchida *et al*.

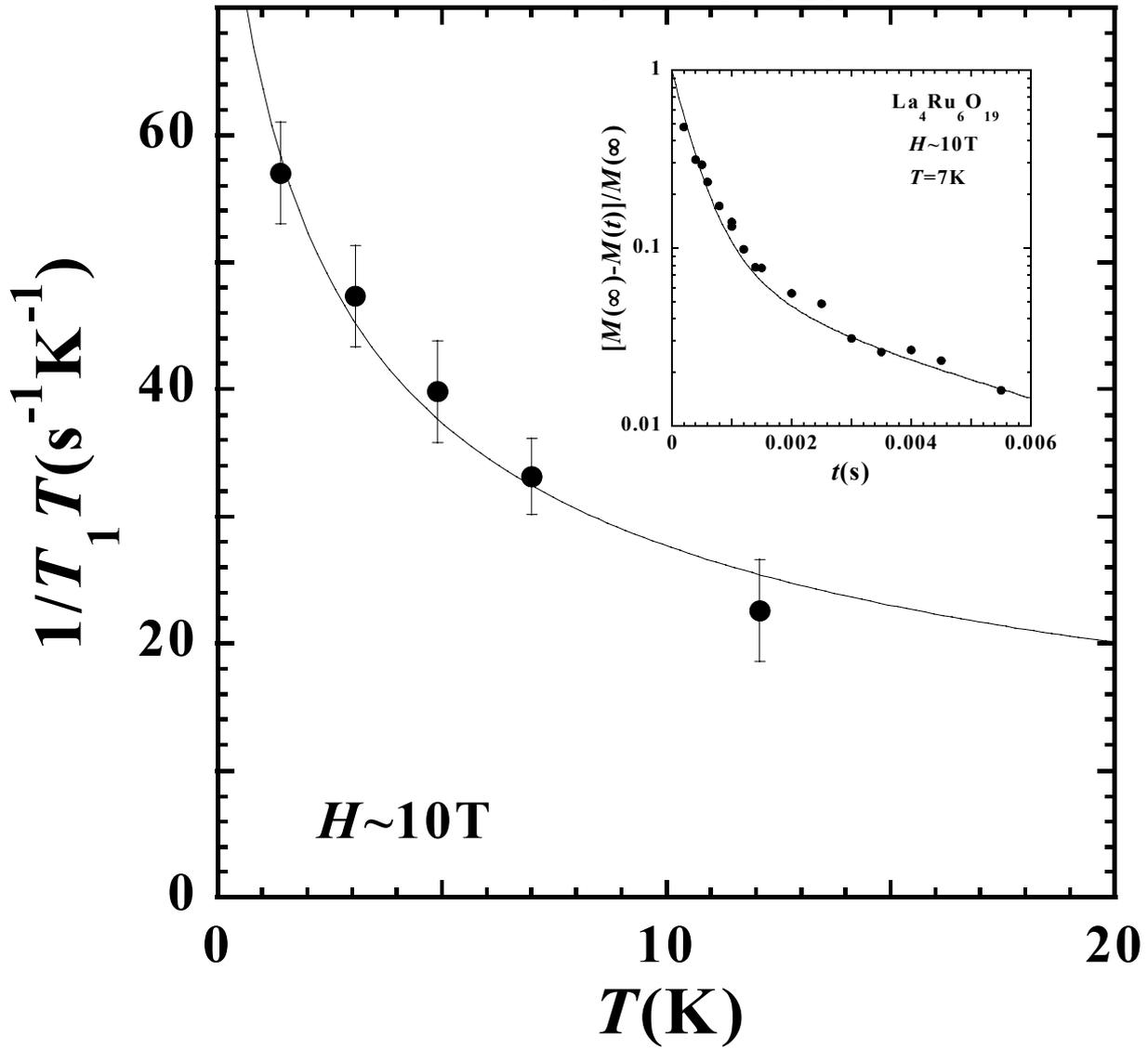

Fig.6
Tsuchida *et al*.

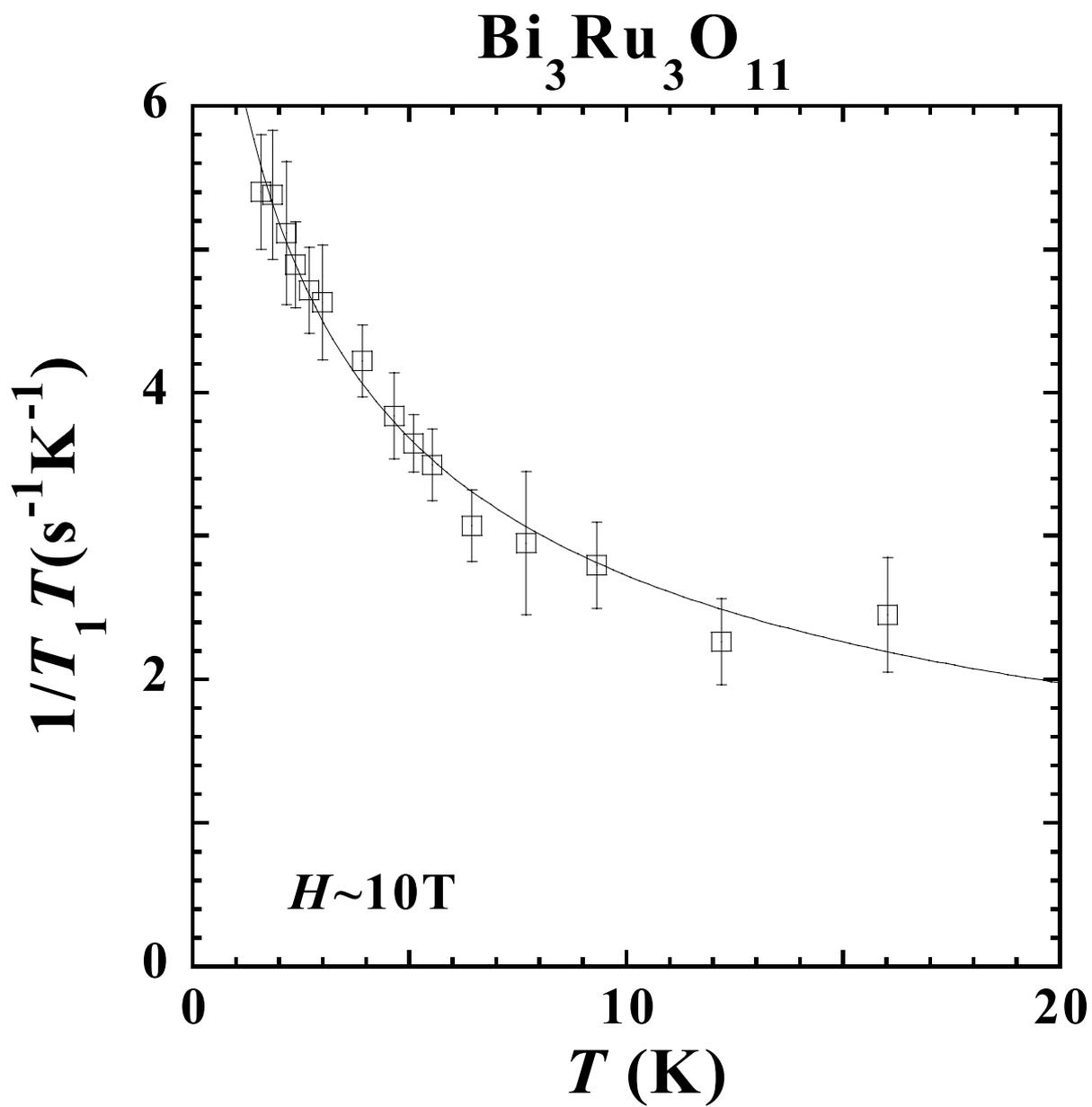

Fig.7
Tsuchida et al.

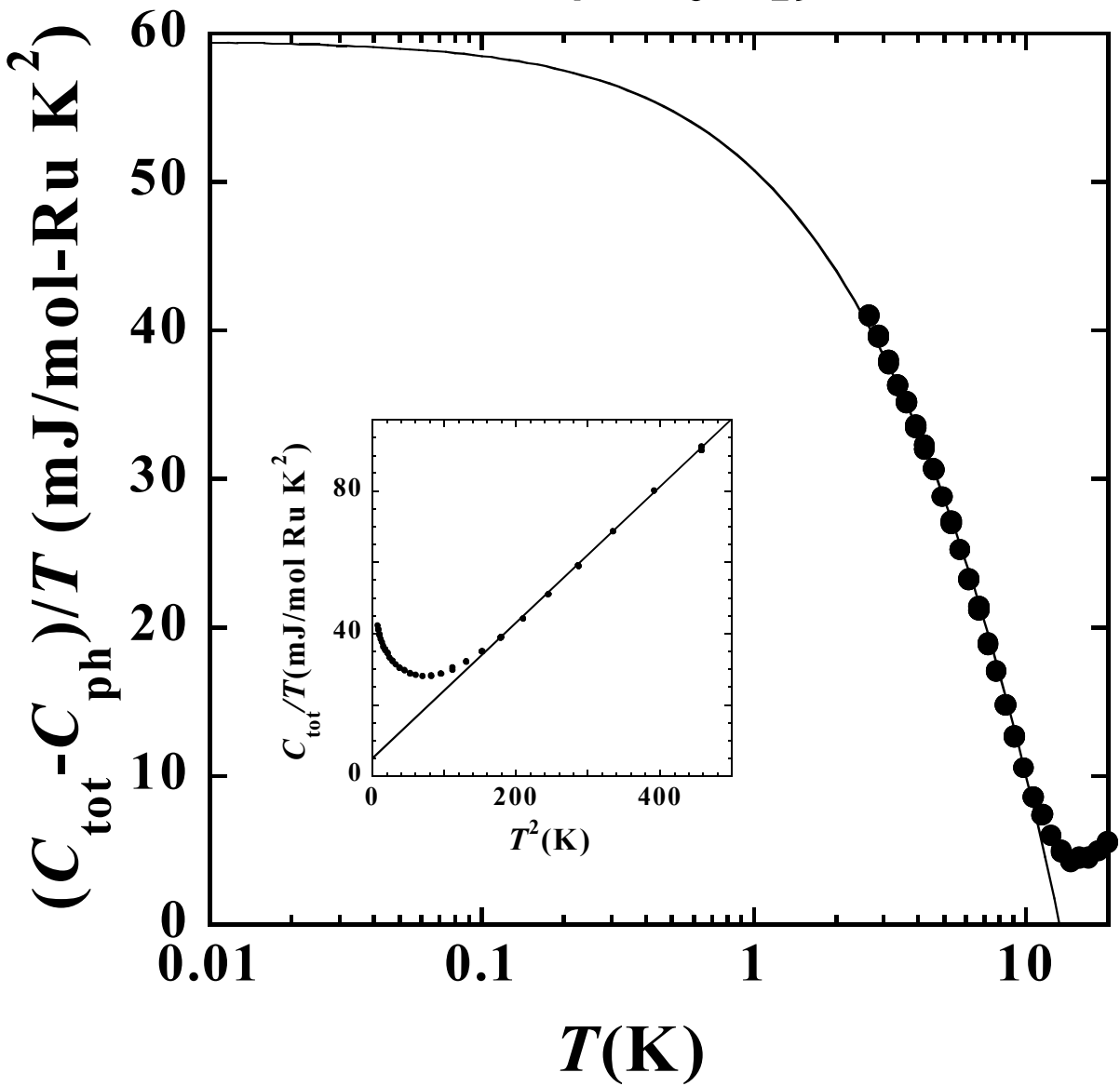

Fig.8
Tsuchida *et al*.